\documentclass[twocolumn,showpacs,preprintnumbers,amsmath,amssymb]{revtex4}
\usepackage{graphicx}
\usepackage{dcolumn}
\usepackage{bm}
\begin{document}

\title{ Enhancement of quantum correlations for the system of cavity QED by applying bang-bang pulses}

\author{Hang-Shi Xu}
\author{Jing-Bo Xu}
 \email{xujb@zju.edu.cn}
\affiliation{Zhejiang Institute of Modern Physics and Physics Department, Zhejiang University, Hangzhou 310027,
People's Republic of China}

\begin{abstract}
We propose a scheme of increasing quantum correlations for the cavity quantum electrodynamics system consisting of two noninteracting two-level atoms each locally interacting with its own quantized field mode by bang-bang pulses. We investigate the influence of the bang-bang pulses on the dynamics of quantum discord, entanglement, quantum mutual information and classical correlation between the two atoms. It is shown that the amount of quantum discord and entanglement of the two atoms can be improved by applying the bang-bang pulses.
\end{abstract}

\pacs{03.67.-a, 03.65.Ud}

\maketitle

\section{Introduction}
Quantum entanglement is at the heart of the current development of quantum information and quantum computation [1]. It is a special quantum correlation and has been recognized as an important resource in quantum information processing [2-4]. The experimental demonstrations of two-particle entanglement and multi-particle entanglement in the cavity quantum electrodynamics (QED) have been reported [5, 6]. Some applications which focus on the entanglement or the nonclassical correlations have also been realized in recent experiments [7-10].

However, the entanglement is not the only type of quantum correlation and there exist quantum tasks that display  quantum advantage without entanglement [11-14]. It has been demonstrated both theoretically [15-17] and experimentally  [18] that other nonclassical correlation, namely, quantum discord [19] can be responsible for the computational speedup for certain quantum tasks. Quantum discord, introduced in [19], is defined as the difference between the quantum mutual information and the classical correlation and is nonzero even for separate mixed states. Therefore, the quantum discord may be regarded as a more general and fundamental resource in quantum information processing.
Recently, the dynamics of entanglement and quantum discord for some open systems has attracted much attention [20-25]. It has been shown that the quantum discord can be completely unaffected by certain decoherence environment during an initial time interval [22] and this phenomenon has been verified by the recent experiment [23].

The interaction between the environment and quantum system of interest can destroy quantum coherence and lead to decoherence. It is therefore of great importance to prevent or minimize the influence of environmental noise in the practical realization of quantum information processing. One of the protocols to prevent the quantum decoherence is dynamical decoupling strategies [26-28] by means of a train of instantaneous pulses("bang-bang" pulses). Recently, experimental suppression of polarization decoherence in a ring cavity using bang-bang decoupling technique has also been reported [29].

In this Letter, we propose a scheme of increasing quantum correlations for the cavity quantum electrodynamics system consisting of two noninteracting two-level atoms interacting with their own quantized field mode [25] by means of a train of instantaneous pulses. The two atoms are initially prepared in the extended Werner-like states(EWL) [30] and the cavity fields are prepared in the Fock states or thermal states. We investigate how the bang-bang pulses affect the dynamics of quantum discord, entanglement, quantum mutual information and classical correlation between the two atoms. It is found that the amount of quantum discord and entanglement of the two atom can be improved by applying the bang-bang pulses, because the increased amount of quantum mutual information is greater than classical correlation by the bang-bang pulses.

\section{Dynamical evolution for the system of cavity QED with bang-bang pulses}
In this section we investigate the dynamical evolution for the cavity quantum electrodynamics system consisting of two noninteracting two-level atoms each locally interacting with its own quantized field mode with bang-bang pulses. The Hamiltonian of one atom interacting with its own quantized field mode with bang-bang pulses is given by
\begin{equation}
H=H_0+H_I+H_P
\end{equation}
with
\begin{equation}
H_0=\frac{\omega_0}{2}\sigma_z+\omega a^\dag a; \quad
H_I=g(\sigma_+a+\sigma_-a^\dag),
\end{equation}
where $a$ and $a^\dag$ denote the annihilation and creation
operators for the cavity field and $\sigma_z=|e\rangle\langle
e|-|g\rangle\langle g|$, $\sigma_+=|e\rangle\langle g|$,
$\sigma_-=|g\rangle\langle e|$ are the atomic operators. $H_P$ is the
Hamiltonian for a train of identical pulses of duration $\tau$, i.e.,
\begin{equation}
H_P=V\sigma_z\sum_{n=0}^\infty \theta(t-T-n(T+\tau))\theta((n+1)(T+\tau)-t),
\end{equation}
where  $T$ is the time interval between two consecutive pulses and the amplitude $V$ of the control
field is specified to be $\frac{\pi}{2\tau}$, which means that we consider the $\pi$-pulse only.

It is not difficult to write down the time evolution operator in the absence of control pulses field
directly as $U_0(T)=exp[-i(H_0+H_I)T]$. With the help of an $SU(2)$ dynamical algebraic structure [31], we
can rewrite the time evolution operator as
\begin{eqnarray}
U_0(T)=\{\cos[\Omega(K)T]-i\frac{\sin[\Omega(K)T]}{\Omega(K)}[\frac{\delta}{2}\sigma_z+\nonumber\\
\quad g(\sigma_+a+\sigma_-a^\dag)]\}\exp[-i\omega(K-\frac{1}{2})T],\qquad
\end{eqnarray}
where $K=\frac{\sigma_z+1}{2}+a^\dag a$ is a constant of motion in
the Hamiltonia, $\delta$ denotes detuning given by
$\delta=\omega_0-\omega$, and
$\Omega(K)=\sqrt{\frac{\delta^2}{4}+g^2K}$. When the control
pulses field is present, the time evolution operator for the duration $\tau$ is given by
\begin{equation}
U_{P}(\tau)=\exp[-i(H_0+H_I+ \frac{\pi}{2\tau}\sigma_z)\tau].
\end{equation}
For the case that the pulses are strong enough, i.e. the duration
$\tau\rightarrow 0$, this time evolution operator reduces to
\begin{equation}
U_{P}\simeq\exp[- i\frac{\pi}{2}\sigma_z],
\end{equation}
which leads to
\begin{equation}
U_{P}U_0(T)U_{P}=-exp[-i(H_0-H_I)T].
\end{equation}

The time evolution operator of an elementary cycle between
$t_{2(N-1)}(=2(N-1)(T+\tau))$ and $t_{2N}(=2N(T+\tau))$ is described by the unitary operator
\begin{equation}
U(t_{2(N-1)},t_{2N})=U_{P}(\tau)U_0(T)U_{P}(\tau)U_0(T)\equiv U_C.
\end{equation}
If we focus on the stroboscopic evolution at times $t_{2N}$,
the evolution is driven by an effective average Hamiltonian [26]
\begin{equation}
U(t_{2N})=[U_C]^N=\exp[-iH_{eff}t_{2N}].
\end{equation}
If $T$ is sufficiently short, then the effective Hamiltonian
is accurately represented by the following Hamiltonian
\begin{equation}
H_{eff}=H_0-ig_{eff}(\sigma_+a-\sigma_-a^\dag).
\end{equation}
The coupling parameter $g_{eff}=\frac{1}{2}g\delta T$ is proportional to the detuning $\delta$ and the time interval $T$ between two successive pulses. Obviously, the interaction between the atom and field is averaged to zero by the "bang-bang" pulses when $T\rightarrow 0$. With the help of the $SU(2)$ algebraic structure as before, the evolution operator at times $t_{2N}$ can be expressed as
\begin{eqnarray}
U(t_{2N})=\{\cos[\Omega_{eff}(K)t_{2N}]-i\frac{\sin[\Omega_{eff}(K)t_{2N}]}{\Omega_{eff}(K)}[\frac{\delta}{2}\sigma_z-\nonumber\\
\quad ig_{eff}(\sigma_+a-\sigma_-a^\dag)]\}\exp[-i\omega(K-\frac{1}{2})t_{2N}],\qquad
\end{eqnarray}
where $\Omega_{eff}(K)=\frac{\delta}{2}\sqrt{1+g^2T^2K}$. The expression for the evolution operator $U(t_{2N})$ in the closed subspace $\{|g\rangle|n\rangle, |e\rangle|n-1\rangle\}$ can be obtained easily from Eq. (11).

In general, at a certain time $t=t_{2N}+\overline{t}$, the evolution operator is given by
\begin{gather}
U(t)=
\begin{cases}
U_0(\overline{t})[U_c]^N & \text{$0\leq\overline{t}<T$}\\
U_0(\overline{t}-T)U_{P}U_0(T)[U_c]^N & \text{$T\leq\overline{t}<2T$},
\end{cases}
\end{gather}
where $N=[\frac{t}{2T}]$, $[\quad]$ denotes the integer part, and the $\overline{t}$ is the residual time after $N$ cycles. Notice that the evolution operators $U_P$ and $U_0(t)$ are closed in the subspace $\{|g\rangle|n\rangle, |e\rangle|n-1\rangle\}$ and the elements of matrixes for $U_P$ and $U_0(t)$ in this subspace can be calculated as
\begin{align}
\langle g|\langle n|U_P|g\rangle|n\rangle=-\langle e|\langle n-1|U_P|e\rangle|n-1\rangle= - i\\
\langle g|\langle n|U_P|e\rangle|n-1\rangle=\langle e|\langle n-1|U_P|g\rangle|n\rangle=0,
\end{align}
and
\begin{eqnarray}
\langle g|\langle n|U_0(t)|g\rangle|n\rangle=\{\cos[\Omega(n)t]+\nonumber\\
i\frac{\delta}{2\Omega(n)}\sin[\Omega(n)t]\}\exp[-i\omega(n-\frac{1}{2})t]\\
\langle e|\langle n-1|U_0(t)|e\rangle|n-1\rangle=\{\cos[\Omega(n)t]-\nonumber\\
i\frac{\delta}{2\Omega(n)}\sin[\Omega(n)t]\}\exp[-i\omega(n-\frac{1}{2})t]\\
\langle g|\langle n|U_0(t)|e\rangle|n-1\rangle=\langle e|\langle n-1|U_0(t)|g\rangle|n\rangle\nonumber\\
=-ig\sqrt{n}\frac{\sin[\Omega(n)
t]}{\Omega(n)}\exp[-i\omega(n-\frac{1}{2})t].
\end{eqnarray}
Then, the explicit expression for the evolution operator $U(t)$ in this subspace can be obtained from Eqs. (12)-(17).

We assume that the two cavity fields are initially in the thermal state
$\sum^\infty_{n=0}p_{1n}|n\rangle\langle
n|\otimes\sum^\infty_{n=0}p_{2n}|n\rangle\langle
n|$, where $p_{in}=\frac{\overline{m}_i^n}{(1+\overline{m}_i)^{n+1}}(i=1,2)$, and $\overline{m}_i=\frac{1}{e^{\beta_i\omega}-1}$ is the mean photon number at the inverse
temperature $\beta_i$. The two atoms are initially in the extended Werner-like states defined by
\begin{eqnarray}
\rho_\Phi=a|\Phi\rangle\langle\Phi|+\frac{1-a}{4}I;\nonumber\\
|\Phi\rangle=\mu|g\rangle_1|e\rangle_2+\nu|e\rangle_1|g\rangle_2;\nonumber\\
\end{eqnarray}
where $a$ is a real number which indicates the purity of initial states, $I$ is a $4\times 4$ identity matrix, $\mu$ and $\nu$ are complex numbers with $|\mu|^2 +|\nu|^2 = 1$. The reduced density operator of the two atoms at time $t$ can be derived by tracing out the cavity fields,
\begin{eqnarray}
\rho(t)=tr_f\{U^{(1)}(t)U^{(2)}(t)|\rho_\Phi\otimes\sum_{n,m=0}^\infty p_{1n}p_{2m}|n\rangle_{11}\langle
n|\nonumber\\
\otimes |m\rangle_{22}\langle
m|U^{(1)\dag}(t)U^{(2)\dag}(t)\},
\end{eqnarray}
where $U^{(i)}(t)(i=1,2)$ is the evolution operator acting on the
$ith$ atom and cavity field. In the standard product basis
$\{|1\rangle_a\equiv|e\rangle_1|e\rangle_2,
|2\rangle_a\equiv|e\rangle_1|g\rangle_2,
|3\rangle_a\equiv|g\rangle_1|e\rangle_2,
|4\rangle_a\equiv|g\rangle_1|g\rangle_2\}$, the reduced density operator of the two atoms can be calculated as follows,
\begin{align}
\rho(t)=\sum_{k,l=1}^4 \rho_{kl}(t)|k\rangle_{aa}\langle l|
\end{align}
where
\begin{eqnarray}
\rho_{11}(t)=\sum_{n,m=0}^\infty p_{1n}p_{2m} [|f_{en}f_{em}|^2\rho_{11}(0)+|f_{en}h_{em-1}|^2\nonumber\\
\rho_{22}(0)+|h_{en-1}f_{em}|^2\rho_{33}(0)+|h_{en-1}h_{em-1}|^2\rho_{44}(0)];\nonumber\\
\rho_{22}(t)=\sum_{n,m=0}^\infty p_{1n}p_{2m} [|f_{en}f_{gm+1}|^2\rho_{11}(0)+|f_{en}h_{gm}|^2\nonumber\\
\rho_{22}(0)+|h_{gn}f_{em}|^2\rho_{33}(0)+|h_{gn}h_{em-1}|^2\rho_{44}(0)];\nonumber\\
\rho_{33}(t)=\sum_{n,m=0}^\infty p_{1n}p_{2m} [|f_{gn+1}f_{em}|^2\rho_{11}(0)+|f_{gn+1}\nonumber\\
h_{em-1}|^2\rho_{22}(0)+|h_{gn}f_{em}|^2\rho_{33}(0)+|h_{gn}h_{em-1}|^2\rho_{44}(0)];\nonumber\\
\rho_{44}(t)=1-\rho_{11}(t)-\rho_{22}(t)-\rho_{33}(t);\qquad\nonumber\\
\rho_{23}(t)=\sum_{n,m=0}^\infty p_{1n}p_{2m}f_{en}h_{gm}h_{gn}^*f_{em}^*\rho_{23}(0);\quad\\
\rho_{12}(t)=\rho_{13}(t)=\rho_{14}(t)=\rho_{24}(t)=\rho_{34}(t)=0;\qquad\nonumber\\
\rho_{ij}(t)=\rho_{ji}(t)^* \qquad (i,j=1,2,3,4),\qquad\nonumber
\end{eqnarray}
with
\begin{eqnarray}
h_{gn}=\langle g|\langle n|U(t)|g\rangle|n\rangle; \nonumber\\
h_{en-1}=\langle e|\langle n-1|U(t)|g\rangle|n\rangle;(n\geq 1) \nonumber\\
h_{e-1}=0;\nonumber\\
f_{gn+1}=\langle g|\langle n+1|U(t)|e\rangle|n\rangle;\nonumber\\
f_{en}=\langle e|\langle n|U(t)|e\rangle|n\rangle.\nonumber
\end{eqnarray}
Here $\rho_{ij}(0)$($i, j$=$1, 2, 3, 4$) are given by Eq. (18) with
$\rho_{11}(0)=\rho_{44}(0)=\frac{1-a}{4}$, $\quad\rho_{22}(0)=\frac{1+(4|\nu|^2-1)a}{4}$, $\quad\rho_{33}(0)=\frac{1+(4|\mu|^2-1)a}{4}$, $\quad\rho_{23}(0)=\rho_{32}(0)^*=\nu\mu^*a.$
If the two cavity fields are initially in the Fock states $|k\rangle_{11}\langle
k|\otimes |l\rangle_{22}\langle l|$, the reduced density operator of the two atoms at time $t$ has the same form of
Eqs. (20)-(21) provided that the substitutions $p_{1n}\rightarrow\delta_{n,k}$ and
$p_{2m}\rightarrow\delta_{m,l}$ are made.

\section{The influence of bang-bang pulses on the dynamics of quantum correlations}

In this section, we investigate the influence of the bang-bang pulses on the dynamics of quantum correlations for the cavity quantum electrodynamics system consisting of two noninteracting two-level atoms each locally interacting with its own quantized field mode. The definition of quantum discord is based on quantum mutual information which contains both classical and quantum correlations. For a bipartite system $\rho^{AB}$, its total correlations can be measured by its quantum mutual information [19]
\begin{equation}
I(\rho^{AB})=S(\rho^{A}) + S(\rho^{B}) -
S(\rho^{AB}),
\end{equation}
where $S(\rho)=-Tr(\rho\log_2\rho)$ is the von Neumann entropy, and $\rho^{A}$ and $\rho^{B}$ denote the reduced density matrices of parts $A$ and $B$, respectively. The quantum discord is defined
as the difference between the quantum mutual information and the
classical correlation [19],
\begin{equation}
Q(\rho^{AB}) =I(\rho^{AB}) -
C(\rho^{AB}),
\end{equation}
where $C(\rho^{AB})$ is the classical correlation which depends on the maximal information obtained with measurement on one of the subsystems and can be expressed as [33]
\begin{equation}
C(\rho^{AB})=\max_{\{B_{k}\}}[S(\rho^{A}) -
S(\rho^{AB}|{\{B_{k}\})}],
\end{equation}
where $\{B_{k}\}$ is a complete set of projectors preformed on
subsystem $B$ locally, $S(\rho^{AB}|\{B_{k}\})=\sum_k p_kS(\rho_{k})$
is the quantum conditional entropy,
$\rho_k=1/p_k (I\otimes B_k) \rho^{AB}
(I\otimes B_k)$ is the conditional density operator and
$p_k=\mathrm{tr}_{(AB)}[(I\otimes B_k) \rho^{AB} (I\otimes B_k)]$ is the
probability.

\begin{figure}
\begin{center}
{\includegraphics[width=6cm]{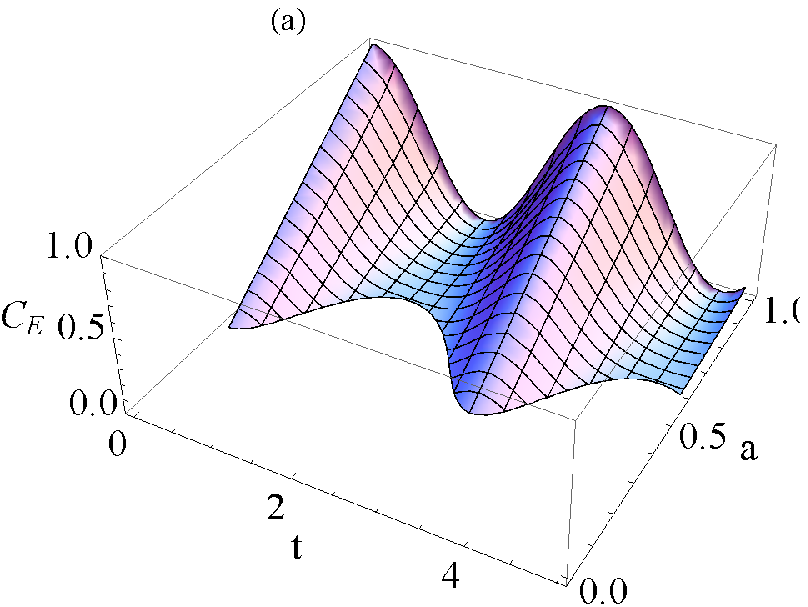}}
\qquad
{\includegraphics[width=6cm]{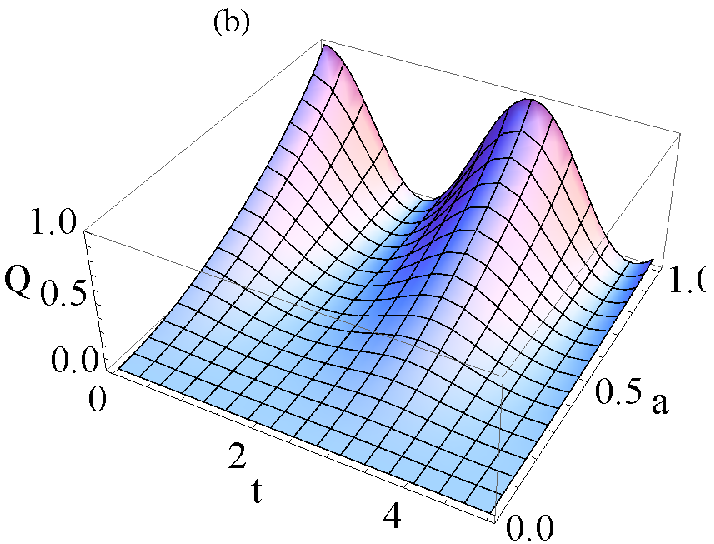}}
\end{center}
\caption{\label{Figure.1 }The concurrence $C_E(t)$(a) and the quantum discord $Q(t)$(b) of two atoms are
plotted as a function of $t$ and $a$ for $g=1, \omega=1, \delta=0,  \overline{m}_1=\overline{m}_2=0$ without control pulses.}
\end{figure}

\begin{figure}
\begin{center}
{\includegraphics[width=8cm,height=4cm]{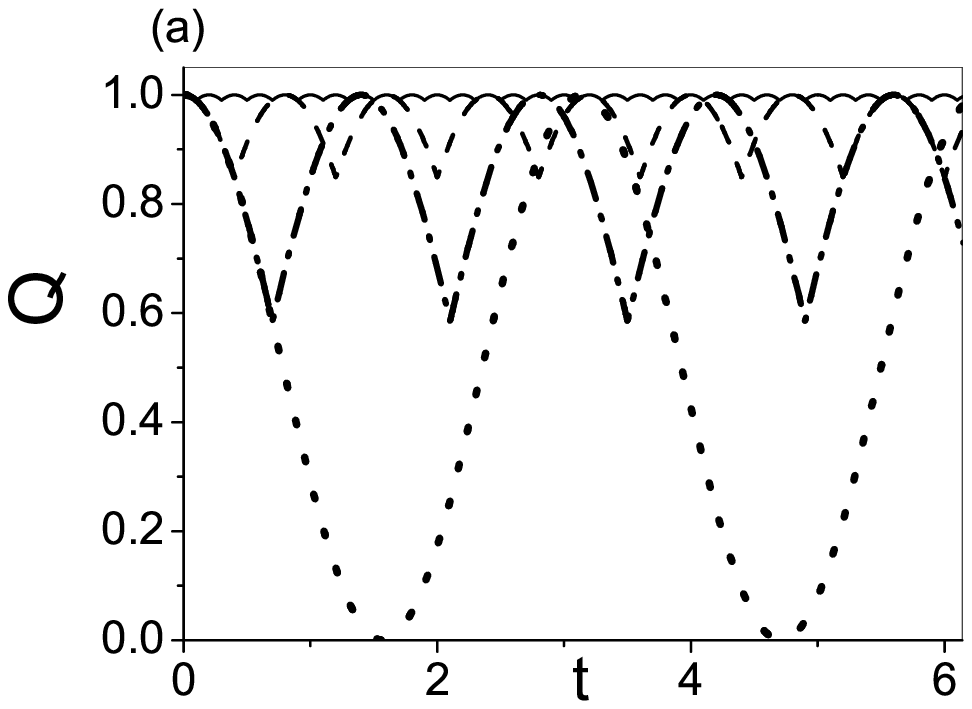}}
\qquad
{\includegraphics[width=8cm,height=4cm]{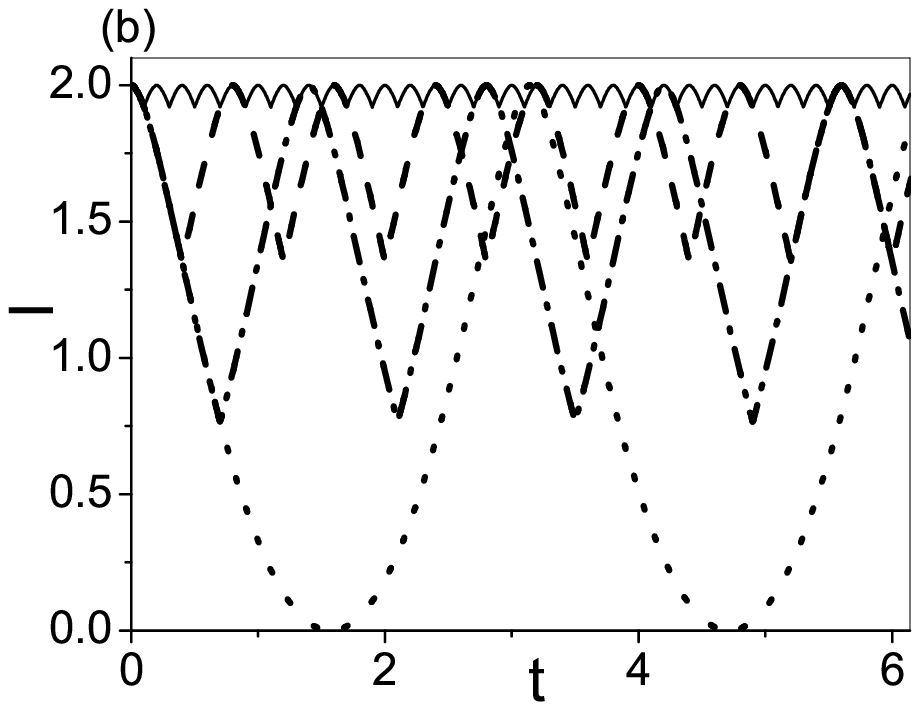}}
\qquad
{\includegraphics[width=8cm,height=4cm]{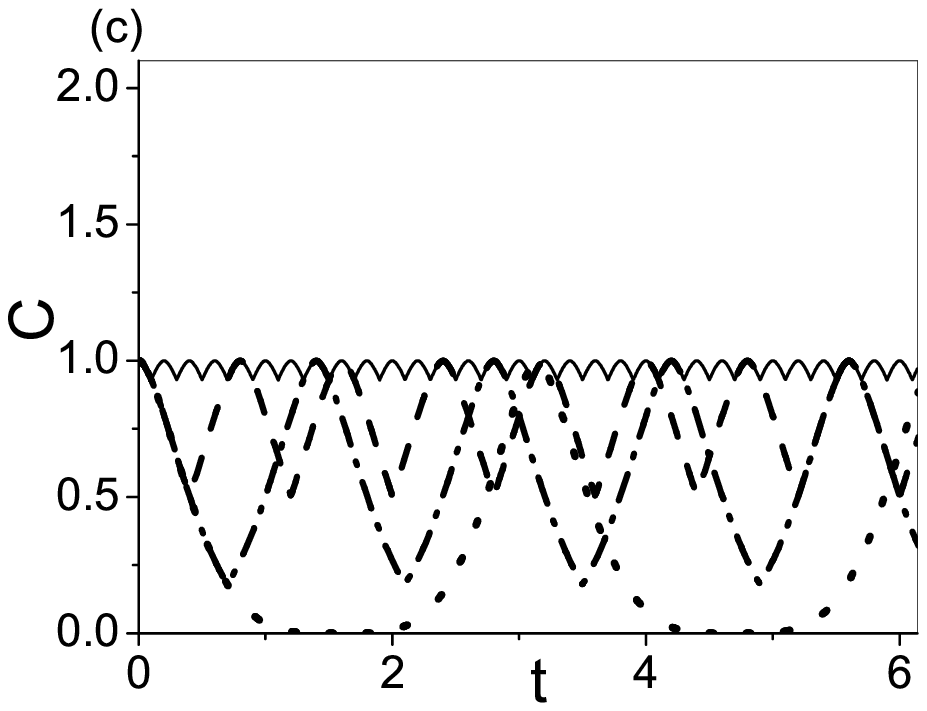}}
\end{center}
\caption{\label{Figure.1 }Quantum discord $Q(t)$(a), Quantum mutual information $I(t)$(b) and classical correlation $C(t)$(c) are plotted as a function of time $t$ with $g=1, \omega=1, \delta=0, a=1,  \overline{m}_1=\overline{m}_2=0$ for different $T$: $T=0.7$(dot-dash line), $T=0.4$(dash line), $T=0.1$(solid line) and without control pulses(dot line).}
\end{figure}

The eigenvalues of the density matrix $\rho(t)$ in Eq. (20) can be calculated as
\begin{align}
\lambda_{1,2}=\frac{1}{2}[(\rho_{11}+\rho_{44})\pm\sqrt{(\rho_{11}-\rho_{44})^2+4|\rho_{14}|^2}]\\
\lambda_{3,4}=\frac{1}{2}[(\rho_{22}+\rho_{33})\pm\sqrt{(\rho_{22}-\rho_{33})^2+4|\rho_{23}|^2}].
\end{align}
Then the quantum mutual information can be derived as follows,
\begin{align}
I(\rho)=S(\rho^{(1)}) + S(\rho^{(2)})+\sum_{i=1}^4\lambda_i\log_2\lambda_i,
\end{align}
\begin{eqnarray}
S(\rho^{(1)})=-[(\rho_{11}+\rho_{22})\log_2(\rho_{11}+\rho_{22})\nonumber\\
+(\rho_{33}+\rho_{44})\log_2(\rho_{33}+\rho_{44})],\\
S(\rho^{(2)})=-[(\rho_{11}+\rho_{33})\log_2(\rho_{11}+\rho_{33})\nonumber\\
+(\rho_{22}+\rho_{44})\log_2(\rho_{22}+\rho_{44})],
\end{eqnarray}
where $\rho^{(i)}(i=1,2)$ is the reduced density matrices of the ith atom. For a two-qubit $X$ state, the quantum discord $Q(\rho)$ between two atoms can be obtained [33, 34].

In order to quantify the entanglement dynamics of the two atoms and make a comparison with the quantum discord dynamics, we use the Wootters concurrence [32] as a entanglement measure. For the state that the density matrix have $X$ structure [33] as Eqs. (20)-(21), the explicit expression of concurrence between two atoms is
\begin{align}
C_E(t)=max\{0,2|\rho_{32}|-2\sqrt{\rho_{11}\rho_{44}}\}.
\end{align}

Generally, for mixed quantum state, the quantum discord does not coincide with entanglement. However, for the two-qubit density matrix of the form $\rho=(1-\alpha)|00\rangle\langle 00|+\alpha|\Psi_m\rangle\langle \Psi_m|$, where $|\Psi_m\rangle$ is maximally entangled state orthogonal to $|00\rangle$ and $\alpha\in [0,1]$, the concurrence is equal to the quantum discord [35]: $C_E=Q=\alpha$. The state described by the density matrix $\rho(t)$(with $a=1,\mu=\pm\nu=\frac{\sqrt{2}}{2}$, $\overline{m}_1=\overline{m}_2=0$) is the concrete situation of this density matrix with
\begin{align}
\alpha=|h_{g0}f_{e0}|^2.
\end{align}

\begin{figure}
\begin{center}
{\includegraphics[width=8cm,height=4cm]{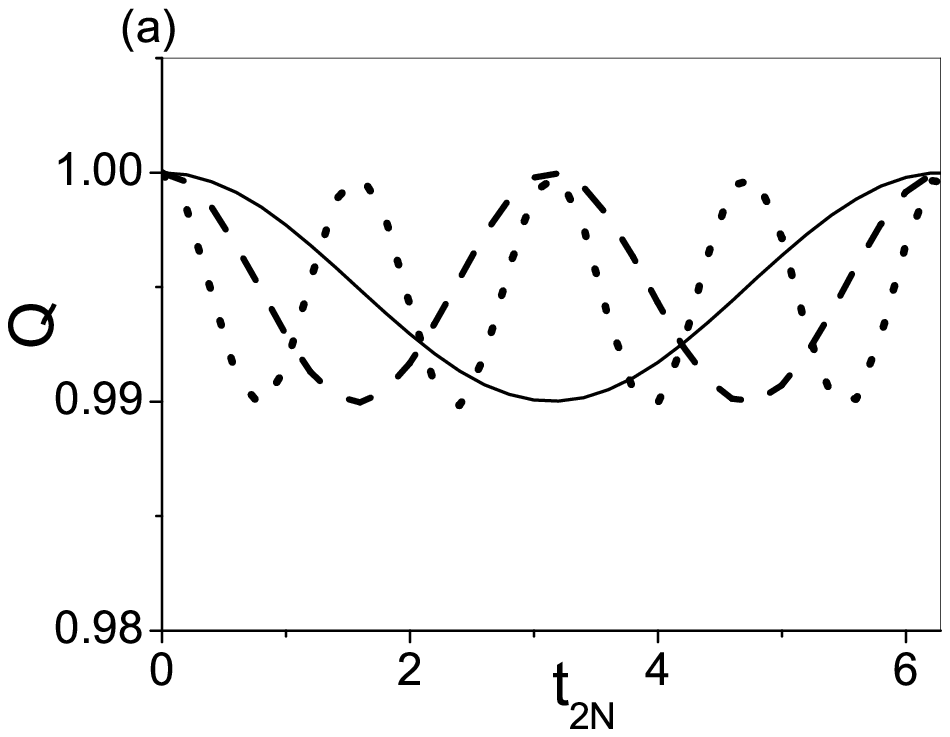}}
\qquad
{\includegraphics[width=8cm,height=4cm]{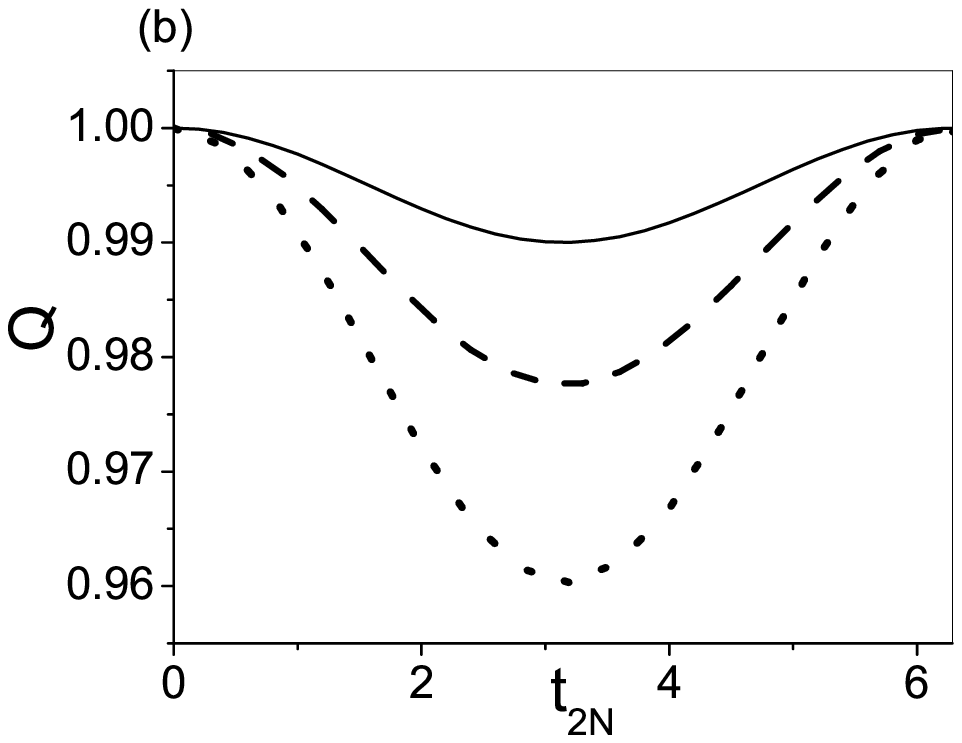}}
\end{center}
\caption{\label{Figure.1 }Quantum discord $Q(t_{2N})$ at the points $t_{2N}$ are plotted as a function of time $t_{2N}$ (a) with $T=0.1, g=1, \omega=1, a=1,  \overline{m}_1=\overline{m}_2=0$ for different $\delta$: $\delta=4$(dot line), $\delta=2$(dash line), $\delta=1$(solid line) and (b) with $\delta=1, g=1, \omega=1, a=1, \overline{m}_1=\overline{m}_2=0$ for different $T$: $T=0.2$(dot line), $T=0.15$(dash line), $T=0.1$(solid line).}
\end{figure}

In Fig. 1, we plot the concurrence and quantum discord for the state described by the density matrix $\rho(t)$ as a function of time $t$ and the parameter $a$ in the absence of control pulses field. The cavity modes are prepared initially in the vacuum states and the two atoms are prepared initially in a Werner state($\mu=-\nu=\frac{\sqrt{2}}{2}$). It can be seen from Fig. 1(a) that the entanglement of two atoms vanishes and revives periodically with time as $a>\frac{1}{3}$ and is always zero for $a\leq\frac{1}{3}$ [33], which means that the entanglement sudden death(ESD) phenomenon [25] appears for the system. The dark period of time, during which the concurrence is zero, is shorter for larger value of $a$. However the quantum discord of two atoms vanishes asymptotically as shown in Fig. 1(b).

\begin{figure}
\begin{center}
{\includegraphics[width=8cm,height=4cm]{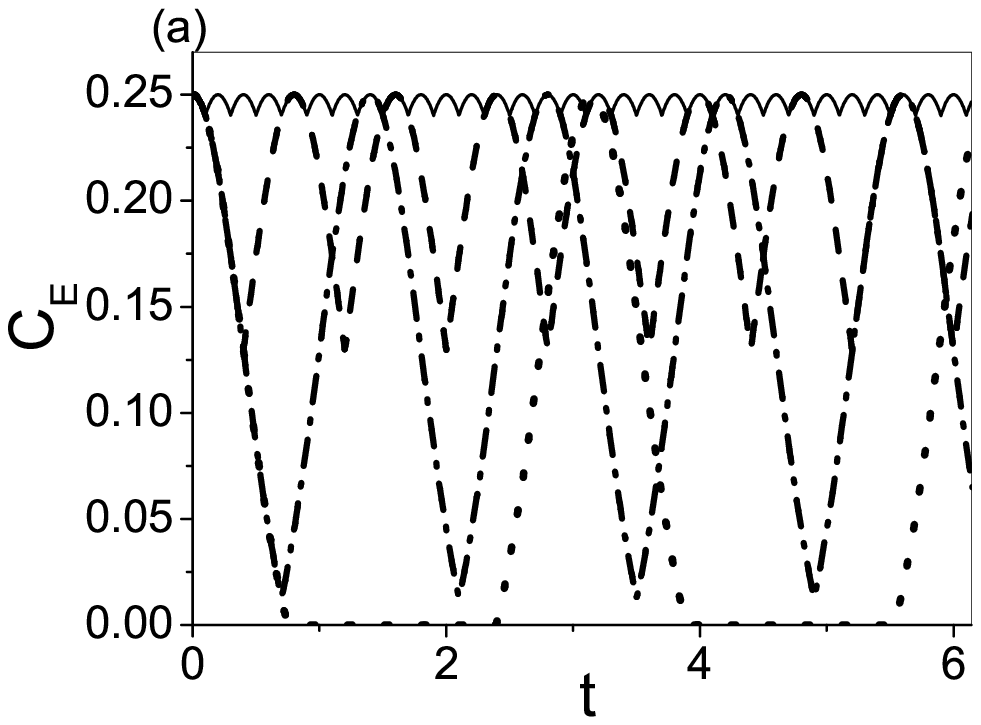}}
\qquad
{\includegraphics[width=8cm,height=4cm]{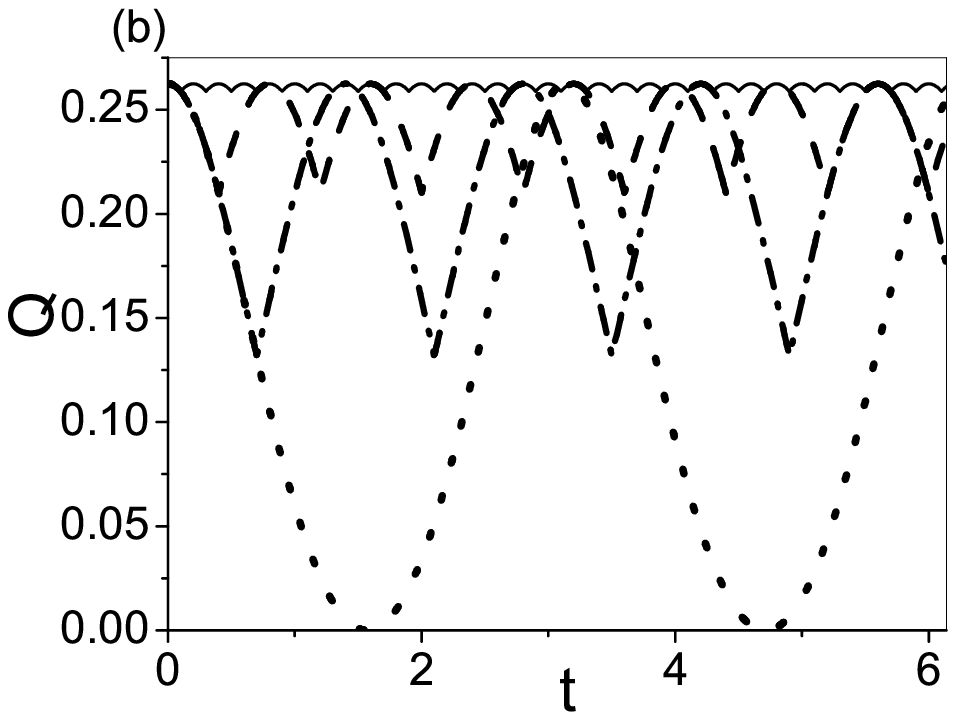}}
\end{center}
\caption{\label{Figure.1 }The concurrence $C_E(t)$(a) and quantum discord $Q(t)$(b)
are plotted as a function of time $t$ with $g=1, \omega=1, \delta=0,  a=\frac{1}{2}, \overline{m}_1=\overline{m}_2=0$ for different $T$: $T=0.7$(dot-dash line), $T=0.4$(dash line), $T=0.1$(solid line) and without control pulses(dot line).}
\end{figure}

\begin{figure}
\begin{center}
{\includegraphics[width=8cm,height=4cm]{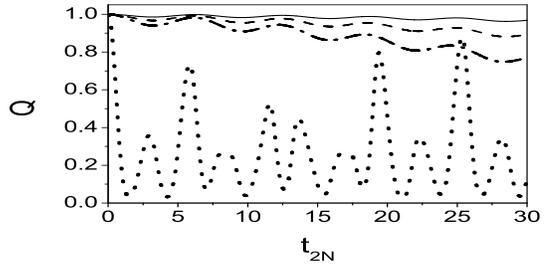}}
\end{center}
\caption{\label{Figure.1 }Quantum discord $Q(t_{2N})$ at the points $t_{2N}$ are plotted as a function of time $t_{2N}$ with $\delta=1, g=1, \omega=1, a=1, \overline{m}_1=\overline{m}_2=0.2$ for different $T$: $T=0.2$(dot-dash line), $T=0.15$(dash line), $T=0.1$(solid line) and without control pulses(dot line).}
\end{figure}

When the bang-bang control pulses field is present, the quantum and classical correlations are displayed as a function of the time in Figs. 2-5. The parameters $a=1$, $\mu=-\nu=\frac{\sqrt{2}}{2}$ are chosen in Figs. 2, 3 and 5, i.e., the two atoms are initially prepared in the maximally entangled state. We can see clearly from Fig. 2 that the quantum discord between the two atoms can be enhanced by the pulses because the increased amount of quantum mutual information is always larger than the classical correlation and the quantum discord recovers to its initial value at the points $t_{2N}$ when the detuning $\delta=0$. The increased amount of the quantum correlations is larger for shorter time intervals $T$ of the control pulses. It is worth pointing out that different choices of $\mu$ and $\nu$ do not give dynamics of quantum correlations qualitatively different from the case treated here.

Focusing on the evolution at times $t_{2N}$, the quantum discord fluctuates with period  $\frac{2\pi}{\delta}$ for the case of the detuning $\delta\neq0$, as displayed in Fig. 3(a) and the amplitudes are independent from the detuning $\delta$. When the cavity fields are prepared in the Fock states, similar results are found for sufficiently short time intervals between two consecutive pulses. These phenomenon can be understood for both $\Omega_{eff}$ and $g_{eff}$ in Eq. (11) are proportional to $\delta$ and $\Omega_{eff}(n)$ is not sensitive to $n$ when $T$ is small. It is quite clear from Fig. 3(b) that the amplitude of quantum discord between two atoms is smaller for pulses with shorter time intervals $T$. In fact, from Eqs. (11), (21) and (31), the quantum discord displayed in Fig. 3 can be given by $Q(t_{2N})\simeq 1-g^2T^2\sin^2\frac{\delta t_{2N}}{2}$ for small $T$. This expression is in accord with Fig. 3.

For the Werner state with $a=\frac{1}{2}$, the concurrence and quantum discord  are plotted in Fig. 4 as a function of time $t$ for different intervals of control pulses. We find that both the concurrence and quantum discord can be enhanced by applying the bang-bang control pulses. It is interesting to point out that the phenomenon of ESD may disappear if the time interval $T$ of the control pulses is sufficiently short.

When the cavity modes are prepared initially in the thermal states, the quantum discord between the two atoms can also be enhanced by the pulses with short time interval as shown in Fig. 5. The changes of quantum discord with different time intervals of the control pulses are synchronous as the same period as the Fig. 3(b). However, different from the case of vacuum states, the maximums of quantum discord are decreasing slightly with time. They decrease more slowly for shorter time intervals $T$ of the control pulses.

\section{Conclusions}
In this paper, we propose a scheme of increasing quantum correlations for the cavity quantum electrodynamics system consisting of two noninteracting two-level atoms each locally interacting with its own quantized field mode by making use of bang-bang pulses. The two atoms are initially prepared in the EWL states and the cavity fields are prepared in the Fock states or thermal states. It is found that the amount of quantum discord and entanglement of two atom can be enhanced by applying the bang-bang pulses and the increased amount is larger for shorter time intervals of the control pulses. Particularly, the phenomenon of ESD may disappear if the time interval $T$ of the control pulses is sufficiently short. In addition, the quantum correlations recover to their initial values at the points $t_{2N}$ when the detuning $\delta=0$. The values of quantum discord at times $t_{2N}$ fluctuates with period $\frac{2\pi}{\delta}$ for the case of the detuning $\delta\neq0$ and the amplitude is smaller for the pulses with shorter time intervals $T$. It is worth noting that the pulses used in this paper are also suitable to the system with the Hamiltonian of interaction has the form of $\sigma_+\hat{B}+\sigma_-\hat{B}^\dag$(here $\hat{B}$ and $\hat{B}^\dag$ are any operators of the cavity field or reservoir). The approach adopted here may be used to improve the implementation of tasks based on quantum correlations in quantum information processing.

\acknowledgments
This project was supported by the National Natural Science
Foundation of China (Grant No. 10774131).

\end{document}